\documentclass[letterpaper,twocolumn,showpacs,prl]{revtex4} % compact publication style
\usepackage{times}
\usepackage{graphics}
\usepackage{graphicx}
\usepackage{epsfig}
\usepackage{amsmath}
\begin{document}
%------------------------------------------------------------------------------------------
%Header

\title{Continuous and Pulsed Quantum Zeno Effect}
\author{Erik W. Streed$^{1,2}$, Jongchul Mun$^{1}$, Micah Boyd$^{1}$, Gretchen K. Campbell$^{1}$, Patrick Medley$^{1}$, Wolfgang Ketterle$^{1}$, David E. Pritchard$^{1}$}

\affiliation{ $^{1}$Department of Physics, MIT-Harvard Center for Ultracold Atoms, and Research Laboratory of Electronics, MIT, Cambridge, Massachusetts 02139, USA\\
$^{2}$Centre for Quantum Dynamics, Griffith University, Nathan, QLD
4111, Australia }

\date{\today}

\begin{abstract}
Continuous and pulsed quantum Zeno effects were observed using a
$^{87}$Rb Bose-Einstein condensate(BEC). Oscillations between two
ground hyperfine states of a magnetically trapped condensate,
externally driven at a transition rate $\omega_R$, were suppressed
by destructively measuring the population in one of the states with
resonant light. The suppression of the transition rate in the two
level system was quantified for pulsed measurements with a time
interval $\delta t$ between pulses and continuous measurements with
a scattering rate $\gamma$. We observe that the
 continuous measurements exhibit the same suppression in the
transition rate as the pulsed measurements when $\gamma\delta
t=3.60(0.43)$, in agreement with the predicted value of 4.
Increasing the measurement rate suppressed the transition rate down
to $0.005\omega_R$.
\end{abstract}

\maketitle

The quantum Zeno effect (QZE) is the suppression of transitions
between quantum states by frequent measurements. It was first
considered as a theoretical problem where the continuous observation
of an unstable particle would prevent its decay \cite{Misra-77}.
Experimental demonstrations of the QZE
\cite{Itano-90,Nagels-97,Molhave-00,Fischer-01,Nakanishi-01,Balzer-02,Kwiat-06}
have been driven by interest in both fundamental physics and
practical applications. Practical applications of the QZE include
reducing decoherence in quantum computing
\cite{Franson-04,Facchi-05,Kwiat-06}, efficient preservation of spin
polarized gases \cite{Nagels-97,Molhave-00,Nakanishi-01}, and dosage
reduction in neutron tomography \cite{Facchi-02}.

The QZE is a paradigm and test bed for quantum measurement
theory\cite{Koshino-05,Home-97}. In one interpretation, it involves
many sequential collapses of the wavefunctions of the system.
Quantum Zeno experiments provide constraints for speculative
extensions of quantum mechanics where the collapse of the
wavefunction is created by extra terms in a modified Schr{\"o}dinger
equation \cite{Adler-03}. It is still an open question how close one
can approach the limit of an infinite number of interrogations due
to the Heisenberg uncertainty involved in shorter measurement times.
These conceptional questions provide the motivation to extend
experimental tests of the quantum Zeno phenonmenon. A major
improvement to a quatum Zeno experiment with ultracold neutrons
\cite{Jaekel-05} is in preparation.

In this letter we compare the suppression of the transition rate in
an oscillating two level system by continuous and pulsed
measurements. Our QZE experiments were carried out with
Bose-Einstein condensed atoms\cite{Anderson-95,Davis-95,Bradley-95}.
The long coherence time and the high degree of control of the
position and momentum of the atoms created a very clean system and
allowed us to observe much stronger quantum Zeno suppression than
before\cite{Itano-90,Fischer-01,Balzer-02}. In the experiment with
pulsed measurements up to 500 measurements could be carried out and
survival probabilities exceeded 98$\%$. Furthermore, we have
performed the first quantitative comparison between the pulsed and
continuous measurement QZE. This is important since any real pulsed
measurement is only an approximation based on a series of weak
continuous measurements \cite{Burn-02,Caves-87}.

Let us consider a two-level system which is externally driven at a
 Rabi frequency $\omega_R$. Measurements of the state of the system project the system
into one of the two states $\left| 1\right>$, $\left| 2\right>$. If
the initial state of the system is in $\left| 1\right>$ and a
measurement is made after short time $\delta t$ ( $ \ll 1/
\omega_R$), then the probability that the system is in $\left|
1\right>$ is $ 1 - ( \omega_R \delta t /2)^2$. With $N$ successive
measurements the probability that the system remains in $\left|
1\right>$ is
\begin{eqnarray}
P \left( N \right) & = & \left[1 - \left( \omega_R \delta t
/2\right)^2 \right]^N  \approx \textrm{exp} \left[ - N \left(
\omega_R \delta t /2\right)^2 \right] \nonumber\\
& =& \textrm{exp} \left[ -\left( {\omega_R}^2 \delta t /4  \right)T
\right] \label{PulsedProb}
\end{eqnarray}
with $T = N \delta t$ the total free evolution time. Instead of
normal Rabi-type oscillation between two states, the initial state
$\left|1\right>$ decays with an effective decay rate
$1/\tau_{EP}$\cite{Schulman-98}. $1/\tau_{EP}$ is given by
\begin{equation}
1/\tau_{EP}={\omega_R}^2 \delta t /4 \label{pulsedrate}
\end{equation}
The characteristic time  $\tau_{EP}$ for the pulsed QZE is much
longer than the characteristic time $1/\omega_R$ of normal Rabi-type
oscillation . This shows the suppression of transition by the QZE.

For a continuous measurement, the atoms are continuously illuminated
with laser light resonant with the transition energy between state
$\left| 2 \right>$ and another excited state. If atoms are in state
$\left| 2 \right>$, they spontaneously emit a photon at a rate
$\gamma$. Due to the photon recoil, those atoms are removed from the
coherently driven two-level system. The population of state $\left|
1 \right>$ decays with the effective decay rate $1/\tau_{EC}$ which
is given by the optical Bloch equations as
\begin{equation}
1/\tau_{EC} = \omega_R ^2 /\gamma .\label{continuouslifetime}
\end{equation}

In contrast, for measurements with randomly spaced pulses, the
effective decay rate is $1/\tau_{EP}={\omega_R}^2 \left<\delta
t^2\right>/4\left<\delta t\right>$. If the probability for
measurement pulse during a time interval $\delta t$ is $\gamma
\delta t$, $\left< \delta t^2\right> = 2 / \gamma^2$ and $\left<
\delta t\right> =  1/ \gamma$. The effective decay rate for this
case is
\begin{equation}
1/\tau_{EP,random} =  \omega_R^2/2\gamma
\label{EqRandomPulseLifetime}
\end{equation}

or twice the value of Eq. (\ref{continuouslifetime}).

In our study we have determined the lifetimes $\tau_{EP}$,
$\tau_{EC}$ with each type of measurement  and used them to verify
the prediction of  Eq.(\ref{pulsedrate}), (\ref{continuouslifetime})
that pulsed measurements with time interval $\delta t$ produce the
same suppression of decay as continuous measurements with a
scattering rate $\gamma$ when $\gamma\delta t=4$ \cite{Schulman-98}.
In particular, by verifying Eq. (\ref{continuouslifetime}), we show
that the continuous measurement process can not be simulated by a
series of random pulses with a rate $\gamma$.

\begin{figure}
\centering{
\includegraphics[hiresbb=true,width=\columnwidth]{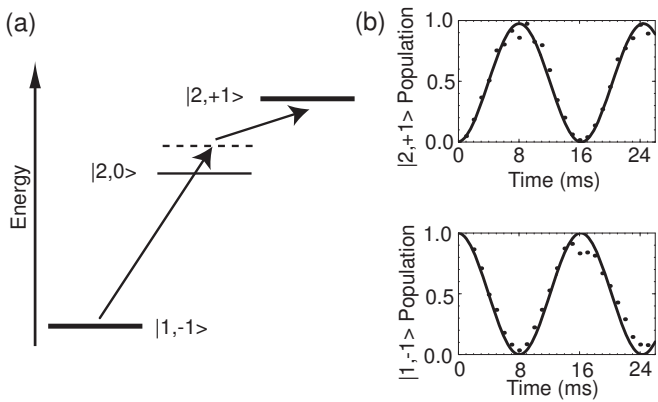}
\caption{ \label{figrabisystemanddetection} Two-level Rabi
oscillation. The two-level quantum system consisted of the
$\left|1,-1\right>$ and $\left|2,+1\right>$ ground hyperfine states
of $^{87}$Rb. a. Energy level diagram for relevant $^{87}$Rb ground
hyperfine states. Arrows depict the components of the two photon
transition between the $\left|1,-1\right>$ and $\left|2,+1\right>$
states. 6.8 GHz microwaves couple the $\left|1,-1\right>$ to a
virtual intermediate state detuned 420 kHz above resonance with
$\left|2,0\right>$. Radio frequency (RF) at 1.68 MHz resonantly
completed the transition to the $\left|2,+1\right>$ state. b. Driven
population of the $\left|1,-1\right>$ and $\left|2,+1\right>$ states
as a function of time. Curves are fits to a two photon transition
rate of $\omega_R/2\pi= 61.5(0.5)$ Hz. No population was detected in
$\left|2,0\right>$.} }
\end{figure}
Our experimental system consisted of magnetically trapped $^{87}$Rb
Bose-Einstein condensate in the 5S$_{1/2}$ $\left|1,-1\right>$
($\left|F,m_F\right>$) and 5S$_{1/2}$ $\left|2,+1\right>$ states.
Pure condensates of $N_c=5.0(0.5)\times10^6$ atoms in the
$\left|1,-1\right>$ state were prepared in a \{63, 63, 6.6\} Hz
magnetic trap \cite{Streed-06}. The atom number was then reduced to
$N_c=5.0(0.5) \times10^4$ by radio frequency(RF) output coupling
\cite{Mewes-97} to lower the density and collisional opacity. The
lifetime of the reduced $\left|1,-1\right>$ condensate exceeded 5 s.
During the experiments a RF shield maintained a magnetic trap depth
of 5 $\mu$K. Coherent oscillations between state
$\left|1\right>(\left|1,-1\right>)$ and state
$\left|2\right>(\left|2,+1\right>)$ were then driven at a rate
$\omega_R$ by a two photon transition (Fig.
\ref{figrabisystemanddetection}). The $\left|1,-1\right>$ and
$\left|2,+1\right>$ states were selected because they have the same
1st order Zeeman shift at a magnetic field of 3.23
G\cite{Harber-02}.

Measurements of the population in  state
$\left|2\right>(\left|2,+1\right>)$ were performed by a laser beam
of 780 nm $\pi$ polarized light resonant with the $5S_{1/2}$
$\left|2,+1\right>$ $\rightarrow$ $5P_{3/2}$ $\left|3,+1\right>$
transition. The laser beam had a $1/e^2$ diameter of $d_0=9.5(0.1)$
mm and its power was monitored with a photodiode. The 362 nK energy
from a single photon recoil distinguished scattered atoms from the
subrecoil $\mu$=15 nK energy range of the condensate atoms.
Successive scatterings would eject measured atoms from the trap.
After each QZE experiment was completed the magnetic trap was turned
off and the population of surviving atoms in each state was
measured. To simultaneously measure the $\left|1\right>$ and
$\left|2\right>$ populations we used an RF pulse and magnetic field
sweep to transfer the atoms to other magnetic sublevels.  Parameters
were chosen in such a way that each initial state was partially
transferred to a sublevel with a different magnetic moment. After
Stern-Gerlach separation and 41 ms of ballistic expansion, the atoms
were imaged and the populations in the two initial states could be
read out simultaneously\cite{Streed-06B}.

\begin{figure}
\centering{
\includegraphics[hiresbb=true,width=\columnwidth]{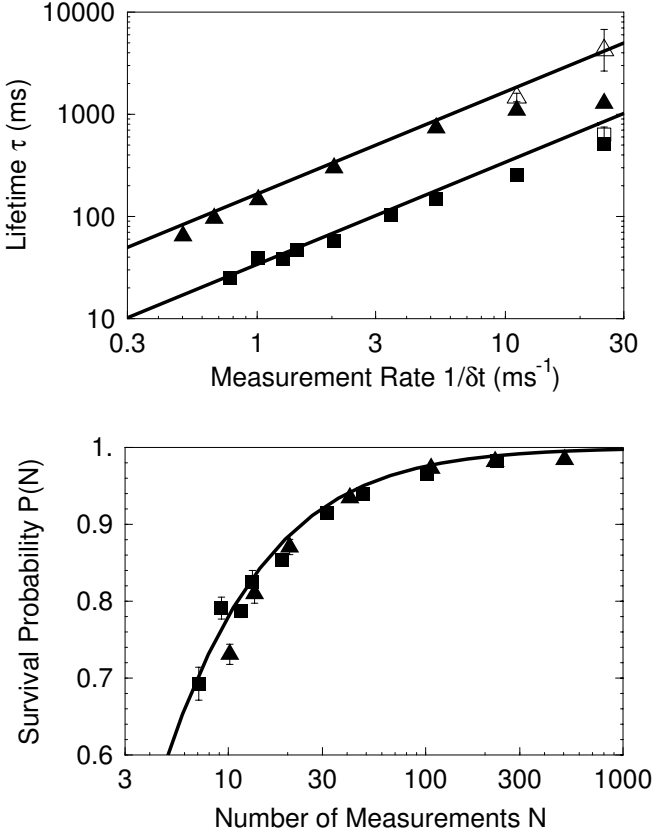}
\caption{\label{figpulsedmeasurements}Pulsed quantum Zeno effect.
Increase in the lifetime (a) and the survival probability (b) of
atoms in the initial $\left|1\right>$ state as the measurement rate
$1/\delta t$ is increased. Solid lines indicate the prediction for
the pulsed QZE. Boxes (triangles) are data points for a transition
rate $\omega_R/2\pi=54.6(0.5)$ ($24.7(0.1)$) Hz. a. Observed
lifetimes (Solid) for $\left|1\right>$ atoms measured with a time
interval $\delta t$ between measurement pulses. Lines indicate the
expected QZE lifetime $\tau_{EP}=4/(\omega_R^2 \delta t)$. Open
symbols show lifetimes after correction for additional loss
mechanism by Eq. \ref{EqPulsedLifetimeCorrection}. b. The same data
is displayed in terms of the survival probability for $N$
measurements performed during a $\pi$ pulse time $t=\pi/\omega_R$ (
the time which would take to  transfer 100\% of the atoms from
$\left|1\right>$ to $\left|2\right>$ without measurements). The
solid line is the expected survival probability $P(N)=[\cos(
\frac{\pi}{2N})]^{(2N)}$ for N ideal measurements.}}
\end{figure}

We quantified the QZE induced by repeated pulsed measurements.
Optical measurement pulses of 172 $\mu$W ($s_0$=0.15, where
$s_0=I/I_{sat}$ is the transition saturation parameter) and $t_p=10$
$\mu$s in duration were applied to the driven two level system. Each
pulse scattered $\sim29$ photons per atom and were separated by a
free evolution time $\delta t$. The lifetime $\tau_{EP}$ for a
particular measurement rate $1/\delta t$ \footnote{Including short
pulse duration $t_p$, the pulse repetition rate in the experiment
was $1/(\delta t + t_p)$} was determined by fitting the
$\left|1\right>$ atom lifetime to an exponential decay curve over a
range of times $ \gtrsim 2\tau_{EP}$. Fig.
\ref{figpulsedmeasurements}a shows the dramatic increase in the
observed lifetimes (solid symbols) as the measurement rate $1/\delta
t$ was increased. The measured lifetimes for two different
$\omega_R$ ( boxes for $2 \pi \cdot 54.6(0.5)$ Hz, triangles for $2
\pi \cdot 24.7(0.1)$ Hz) are plotted along with their expected
values (lower and upper lines respectively). The measured lifetimes
were not found to be strongly sensitive to variations in optical
power, pulse width, or laser detuning. The lifetime enhanced by QZE
can be compared to $1/\omega_R$, which would be the characteristic
time without pulsed measurements. The longest lifetime was $198(16)
\cdot 1/\omega_R$ at $1/\delta t = 25$ms$^{-1}$.

Previous works \cite{Itano-90,Balzer-02,Facchi-03} express the QZE
in terms of the survival probability $P(N)$ for number of
measurements $N$ during a $\pi$ pulse ($t=\pi/\omega_R$), a duration
where without measurements 100\% of the atoms would be transferred
into the other state. Fig. \ref{figpulsedmeasurements}b displays our
results in this way. In these terms the greatest Zeno effect is for
N=506(2) measurements with a survival probability P=0.984(1).

The most frequent measurements (farthest right solid symbols in Fig.
\ref{figpulsedmeasurements}a) show significant deviation from
expected lifetimes(lines). For a high measurement rate $1/\delta t$,
the pulse duration $t_p$ is not negligible compared to free
evolution time $\delta t$ between the pulses and the process that
occurs while the measurement pulse is on becomes more important. In
our experiment the pulse duration $t_p = 10 \mu$s was 20$\%$ of the
shortest time interval $\delta t = 40 \mu$s. In such cases the
measured lifetime depends not only on the time interval $\delta t$
but also on the pulse duration $t_p$. During the time interval,
$\delta t$ the population in state $\left|1\right>$
transfers(``decays'') to state $\left|2\right>$ with $\tau_{EP}$.
During the pulse duration $t_p$, state $\left|1\right>$ can decay by
different loss mechanisms. We made a separate measurement of this
additional loss. The system was prepared in the same way except that
the measurement pulse laser was kept on continuously. The lifetime
$1/\Gamma_m$ of this system was measured and $\Gamma_m$ was
$3.41(0.14) s^{-1}$ for $\omega_R /2\pi = 54.6$Hz $(\Gamma_m = 2.96
(0.22) s^{-1}$ for  $ \omega_R /2\pi = 24.7$Hz $)$. In order to find
the origin of this additional loss, the measurements of lifetimes
were made with removal of either the RF or the microwave component
of the two-photon drive. The lifetime showed no change when the RF
component was removed, but the lifetime increased by an order of
magnitude without microwave component. This suggests that the loss
occuring during pulse duration $t_p$ is dominated by the virtual
intermediate state $\left|2,0\right>$,which can be excited by the
measurement laser to the excited state $5P_{3/2} \left|3,0\right>$

To obtain the correct decay rate $1/\tau_{EP}$ for the pulsed QZE
from our measurement this additional loss should be corrected for.
The observed decay rate $1/\tau$ is split into two components and
can be written as
\begin{equation}
\frac{1}{\tau}= \frac{1}{\tau_{EP}} \frac{\delta t}{t_p+\delta t} +
\Gamma_{m} \frac{t_p}{t_p+\delta t}
\label{EqPulsedLifetimeCorrection}
\end{equation}
where $t_p$ is pulse duration. Data points in Fig.
\ref{figpulsedmeasurements}a where this correction had a significant
impact on the lifetime are indicated by open symbols. The predicted
lifetime is $\tau_{EP}=4/(  \omega_R^2 \delta t )$, slightly larger
than the measured $\tau_{EP}=0.836(0.014) \times4/( \omega_R^2
\delta t)$. The discrepancy is possibly due to collisions between
recoiling atoms and the remaining condensate leading to additional
loss.
\begin{figure}
\centering{
\includegraphics[hiresbb=true,width=\columnwidth]{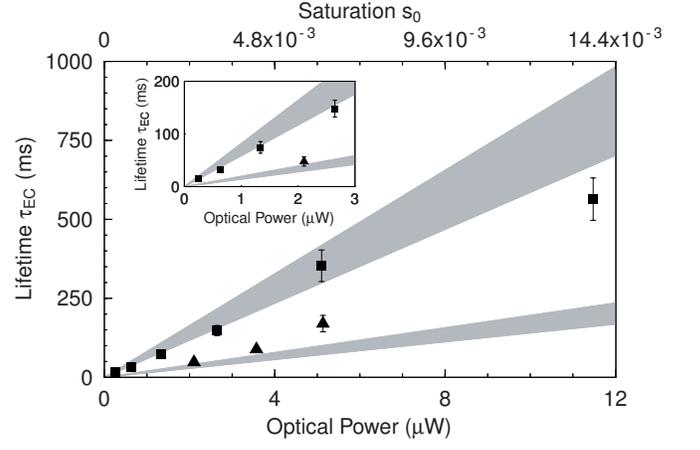}
\caption{\label{figcwlinearities} Continuous quantum Zeno effect.
Lifetime dependence on optical power with $\omega_R/2\pi$ =
$48.5(0.9)$ Hz for laser detuning $\delta_L$ = 0 MHz(boxes) and
$\delta_L$ = $-5.4$ MHz (triangles). Grey bands indicates range of
expected  lifetimes which are calculated from measurements of AC
Stark shift for $\delta_L$ = 0 MHz (upper) and $\delta_L$ = $-5.4$
MHz (lower). Inset highlights data from lower optical powers. The
saturation parameter $s_0$ has an uncertainty of 17\%.} }
\end{figure}

The same initial system was subjected to a weak continuous
measurement instead of repeated strong measurements. Fig.
\ref{figcwlinearities} shows the increase in lifetime with
increasing measurement laser power. While showing this qualitative
relationship is straightforward, several issues complicate a
quantitative measurement of the continuous QZE. If the measurement
laser is detuned from the optical resonance it will have both a
reduced scattering rate and also induce an AC Stark shift
$\delta_{RF}$ in the resonance between $\left| 1 \right>$ and
$\left| 2 \right>$, reducing the effective Rabi frequency. In
addition, imperfections in the beam can affect the intensity at the
atoms. These issues are not important for the pulsed measurement as
long as atoms scatter multiple photons. However they are critical to
properly characterizing the weak continuous measurement experiment.

\begin{figure}
\centering{
\includegraphics[hiresbb=true,width=\columnwidth]{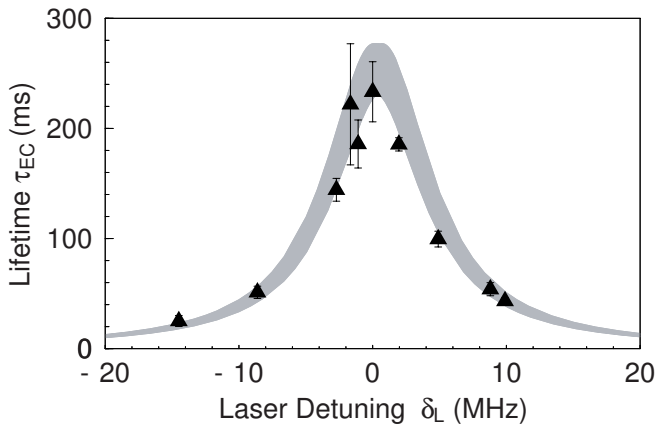}
\caption{\label{figcwscat} Continuous quantum Zeno lifetime as a
function of the measurement laser detuning $\delta_L$. Grey band
indicates range of expected lifetimes(Eq. \ref{EqPredCWLifetime})
from separately measured AC Stark shift parameters. The predicted
linewidth is slightly Zeeman broadened by imperfections in the
polarization. Data is for 3.5 $\mu$W laser power,
$\omega_R=45.5(1.0)$ Hz. } }
\end{figure}

We were able to address all of these issues simultaneously by
measuring the AC Stark shift at several different laser detunings.
For each laser detuning ($\delta_L$) and optical power ($s_0$) we
determined the AC Stark shift $\delta_{RF}$ by maximizing the
reduction of atoms in state $\left| 1 \right>$ as a function of RF
frequency. Measurements of continuous QZE lifetime $\tau_{EC}$ (Eq.
\ref{continuouslifetime}) were then made varying saturation
parameter $s_0$ and detuning $\delta_L$ of the measurement laser.
Eq. \ref{continuouslifetime} can then be rewritten as
\begin{equation}
\tau_{EC}= \frac{ \gamma }{\omega_R^2} =\frac{ \Gamma s_0
}{2\omega_R^2} \left( \frac{1}{1+ 4 \left(\frac{
\delta_L}{\Gamma}\right)^2} \right) \label{EqPredCWLifetime}
\end{equation}
which is a function of $s_0$ and $\delta_L$ with $^{87}$Rb D$_2$
transition decay rate $\Gamma$. Fig. \ref{figcwscat} verifies Eq.
(\ref{EqPredCWLifetime}) for various detunings $\delta_L$. Fig.
\ref{figcwlinearities} shows increasing lifetime with increasing
measurement laser power, the signature of the continuous QZE.
Similar to the longest lifetime point in the pulsed QZE data (upper
right solid triangle, Fig. \ref{figpulsedmeasurements}a), the data
point with highest power in Fig. \ref{figcwlinearities} shows
significant deviation from the lifetime expected from Eq.
\ref{EqPredCWLifetime}. By matching the observed lifetimes for
pulsed and continuous QZE measurements we find that each measurement
type has the same QZE when $\gamma \delta t=3.60(0.43)$, which is in
agreement with the predicted ratio of 4 \cite{Schulman-98} but rules
out randomly repeated pulse case in Eq.
(\ref{EqRandomPulseLifetime}). Eq. (\ref{EqRandomPulseLifetime})
gives the ratio of 2 instead of 4. The observed large quantum Zeno
suppression dramatically illustrates the modification of a
wavefunction by a null measurement, i.e. the observation that no
light has been scattered\cite{Dicke-81}. The large fraction of atoms
in the initial state$\left|1\right>$ is caused by repeated
measurements without scattering any photons.

We have extended previous work in pulsed QZE measurements
\cite{Itano-90,Balzer-02,Fischer-01} by exploiting advantages
inherent to Bose-Einstein condensates. While in theory the
Heisenberg uncertainty principal limits how frequently meaningful
measurements can be performed, in practice imperfections in real
measurements are the limiting factors \cite{Rauch-01,Facchi-03}. In
ion experiments optical pumping between states during the
measurement pulses changed the observed population transfer
\cite{Itano-90}, requiring significant corrections for the N=32 and
N=64 pulse measurements (Table I, \cite{Itano-90}) to observe a
maximum survival probability $P(64)=0.943(20)$ \cite{Itano-90}
($\tau_{EP}=54(30) \cdot 1/\omega_R$). Previous demonstrations of
the continuous QZE \cite{Nagels-97, Molhave-00, Nakanishi-01}
observed qualitative, but not quantitatively characterized QZE
suppression effects up to 80\% \cite{Molhave-00} with increasing
laser intensity. Our observed quantum Zeno suppressions are
substantially larger then both previous pulsed \cite{Itano-90} and
continuous \cite{Molhave-00} results, and is also greater then that
expected from proposed experiments \cite{Rauch-01, Facchi-02,
Facchi-03, Jaekel-05} in neutrons.

In conclusion we have used a Bose-Einstein condensate to demonstrate
the QZE for both continuous and pulsed measurements. Lifetimes for
both cases were substantially enhanced by QZE to values close to
$200\cdot1/\omega_R$ Pulsed and continuous QZE were quantified and
compared. We observe that the continuous measurements exhibit the
same suppression in the transition rate as the pulsed measurements
when $\gamma\delta t=3.60(0.43)$, which agrees with the predicted
value of 4 \cite{Schulman-98} and rules out a simple model when a
continuous measurement is replaced by a series of random pulses. A
next generation experiment could demonstrate even stronger quantum
Zeno suppression and study the transition from pulsed to continuous
QZE by using pulse duration and intervals approaching the
spontaneous emission time.

The authors thank Helmut Rauch for insightful discussion. This work
was supported by NSF.

\end{document}